\documentclass[12pt,showpacs,showkeys,amsmath,aps,prc]{revtex4}

\usepackage{amsfonts} 
\usepackage{mathrsfs}
\usepackage{subfigure}
\usepackage{supertabular}
\usepackage{color,graphicx}
\usepackage[bookmarksopen]{hyperref}

\def  \p    {\pi}

\def  \m    {\mu}
\def  \n    {\nu}

\def  \veps {\varepsilon}

\def  \gf   {\gamma^5}

\def  \del  {\partial}

\def  \sls  {\!\!\!/}

\def  \bef  {\begin{figure}}
\def  \eef  {\end{figure}}
\def  \be   {\begin{equation}}
\def  \ee   {\end{equation}}
\def  \ba   {\begin{array}}
\def  \ea   {\end{array}}
\def  \bea  {\begin{eqnarray}}
\def  \eea  {\end{eqnarray}}
\def  \beq  {\begin{eqnarray}}
\def  \eeq  {\end{eqnarray}}
\def  \nn   {\nonumber}
\def  \bd   {\begin{displaymath}}
\def  \ed   {\end{displaymath}}
\def  \bse  {\begin{subequations}}
\def  \ese  {\end{subequations}}
\def  \bwt  {\begin{widetext}}
\def  \ewt  {\end{widetext}}

\def  \ba   {{\bf{a_1}}}

\begin{document}
\title{Ground state energy of spin polarized quark matter with correlation }
\author {Kausik Pal}
\email {kausik.pal@saha.ac.in}
\author{Subhrajyoti Biswas}
\author {Abhee K. Dutt-Mazumder}
\affiliation {High Energy Physics Division, Saha Institute of Nuclear Physics,
 1/AF Bidhannagar, Kolkata 700064, India.}

\medskip

\begin{abstract}
We calculate the ground state energy of cold and dense spin polarized 
quark matter with corrections due to correlation energy $(E_{corr})$. 
Expressions for $E_{corr}$
both in the non-relativistic and ultra-relativistic regimes have been  
derived and compared with the exchange and kinetic term present
in the perturbation series. It is observed that the inclusion of correlation 
energy does not rule out the possibility of the ferromagnetic phase transition 
at low density within the model proposed by Tatsumi\cite{tatsumi00}. We
also derive the spin stiffness constant in the high density
limit of such a spin polarized matter.
\end{abstract}
\vspace{0.08 cm}



\maketitle

\section{Introduction}

The possibility of ferromagnetic phase transition in dense quark matter 
was first discussed by Tatsumi\cite{tatsumi00} where it was shown that
quark liquid interacting through one gluon exchange shows spontaneous
magnetic instability at low densities. Such an investigation was motivated
by the observation of strong magnetic field in neutron star. Moreover, the
theoretical conjectures about the possible existence 
of quark stars provide additional impetus to examine this issue further
\cite{pal09,niegawa05,tatsumi07,ohnishi07,son08}.

The underlying mechanism of such a phase transition for slow moving 
massive quark is similar to what one observes in case of interacting electron 
gas \cite{gellmann57,pines58} in a neutralizing positive charge 
background where the electron interact 
only by the exchange interaction and the contribution of the direct term 
cancels with the background contribution. In case of interacting electron
gas, the kinetic energy is minimum in unpolarized state, the exchange
energy, on the other hand, favors spin alignment. These are two competing 
phenomenon which also depends on density. It is seen that the kinetic
energy dominates at higher density and as the density is lowered the exchange 
energy becomes larger at some point turning the electron gas suddenly 
into a completely polarized state. This is the mechanism of Ferromagnetism
in electron gas interacting via. Coulomb potential \cite{bloch29}. 

The exchange energy for quark matter interacting via 
one gluon exchange (OGE) is also attractive
and becomes dominant at some density giving rise to Ferromagnetism
\cite{tatsumi00,pal09,niegawa05,tatsumi07}. However,
there are similarities and differences between quark matter and electron
gas as discussed in ref.\cite{tatsumi00}. For slow moving massive
quarks the dynamics is very similar to what happens in electron gas, while
in the relativistic case a completely different mechanism works when
spin dependent lower component of the Dirac spinor becomes important. 
It should also be noted that the exchange energy is negative for massive 
strange quark at low densities while it is always positive for massless u and 
d quark as observed in \cite{chin79prl} and subsequently in
\cite{tatsumi00,tatsumi08}.

The magnetic property of the quark matter was also studied in \cite{niegawa05}
by evaluating the effective potential by employing magnetic moment of a
quark and treating this as an order parameter. Unlike \cite{tatsumi00}, in
this model u, d and s quarks {\em i.e.} all of these flavors, 
show ferromagnetic phase transition at
various densities. In \cite{pal09}, we revisited this problem and have
evaluated Fermi Liquid parameters for a spin polarized quark matter
which were subsequently used to derive single particle spectrum and
total energy density as a function of the 
$\xi=(n_q^+ - n_q^-)/(n_q^+ + n_q^-)$. There it
was shown that such a phase transition within the OGE model and 
parameter set of ref.\cite{tatsumi00}, is possible at very low density.

In \cite{tatsumi00,tatsumi07} and \cite{pal09} calculations were  
restricted only to the Hartree Fock level and the higher order 
terms were ignored.
The computation of the ground state energy on the other hand requires
evaluation of the diagrams beyond the exchange loop {\em viz.} the 
inclusion of correlation energy as emphasized in ref.\cite{tatsumi00}. 
This is rather tricky 
as the higher order terms are plagued by infrared divergences due to
the exchange of massless bosons like gluons (or photons) indicating
the failure of naive perturbation theory. The problem can be handled
by summing a class of diagrams which makes the perturbation series
convergent and receives logarithmic corrections. 
In the case of degenerate electron matter this pioneering 
work was done by Gell-Mann and Brueckner (GB) commonly known as GB
theory where the `correlation energy $(E_{corr})$' of an electron gas 
at high density was calculated \cite{gellmann57}. The correlation energy
is actually the higher order correction to the ground state energy beyond
the exchange term in the perturbation series defined by 
\cite{gellmann57,pines58}

\bea
E_{corr}=E-E_{ex}-E_{kin}
\eea  

Here, $E_{corr}$, $E_{ex}$ and $E_{kin}$ correspond to correlation, 
exchange, kinetic energy density respectively. In general for electron gas interacting via. Coulomb force it takes the following form 
\cite{gellmann57,pines58}

\bea\label{rs_expan}
E_{corr}=A \ln{r_s}+C +{\cal O}(r_s)
\eea

At large Fermi momentum $(p_f)$ {\rm i.e.} in the limit 
$r_s=0$, the result becomes exact \cite{wal_book,pines_book}.
For the case of electron gas, the inverse density
is set equal to $\frac{4}{3}\pi r_0^3$ and the dimensionless parameter $r_s$
is defined as $r_0$ divided by Bohr radius \cite{gellmann57}.
We here, derive a similar expression for the dense quark matter 
with arbitrary spin polarization with appropriate modifications.

The model adopted in the present work is same as that of ref.\cite{tatsumi00}
except here we go beyond ${\cal O}(g^2)$ and include ring diagrams 
to evaluate the correlation energy of spin polarized quark matter.
This, together with the contribution of $E_{kin}$ and $E_{ex}$, 
as we shall see, has the small $\xi$ expansion 

\bea\label{xi_expan}
E &=& E(\xi=0)+\frac{1}{2}\beta_s \xi^2+.....
\eea

Here, $\beta_s=\frac{\del^2 E}{\del\xi^2}{\Big|}_{\xi=0}$, is defined to be 
the spin stiffness constant in analogy with \cite{perdew92,rajiv89}
with $r_s=g^2(\frac{3\pi}{4})^{1/3}$.
It is to be noted that in Eq.(\ref{xi_expan}) first term corresponds to
unpolarized matter with correlation energy having the form of 
Eq.(\ref{rs_expan}). Clearly, this is reminiscent of what one obtains for the 
degenerate electron gas \cite{perdew92}.

The derivation of $E_{corr}$ here requires the evaluation of the gluon 
self-energy in spin asymmetric quark matter which can be used to construct
the in-medium one loop corrected gluon propagator with explicit spin
parameter dependence. Apart from the calculation of correlation energy,
this might have applications in evaluation, for example, of the Fermi
Liquid parameter (FLP) in spin polarized matter or spin susceptibility or
quantities which can be expressed in terms of FLPs 
\cite{pal09,tatsumi08,baym76,matsui81}. 
In the present work,
we however, restrict ourselves to the evaluation of the ring diagrams only.

The plan of the paper is as follows. In Sec. II, we derive the
expression for gluon self energy in polarized quark matter -
an essential ingredient for the calculation of correlation energy.
In Sec. III, we calculate ground state 
energy with correlation correction for the polarized matter. Subsequently,
we also compare exchange and correlation energy density. In Sec. IV 
we summarize and conclude. The detailed expression of various matrix elements 
required to evaluate polarization tensor have been relegated to the Appendix.

\section{Gluon self-energy in polarized matter}

\begin{figure}[htb]
\begin{center}
\resizebox{5.0cm}{3.0cm}{\includegraphics[]{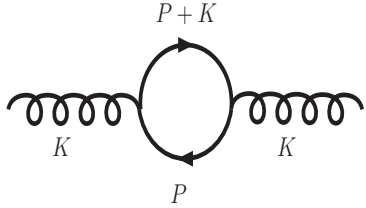}}
\caption{Gluon self energy.}
\label{fig_01}
\end{center}
\end{figure}

To calculate the correlation energy of spin polarized quark matter 
one needs to calculate the gluon self energy in matter with arbitrary
spins. This spin dependent gluon polarization arises
from the quark-loop shown in Fig.(\ref{fig_01})\cite{dumitru09}.
Mathematically \cite{bel_book,niegawa03},

\beq\label{pi_munu}
{\Pi}_{\mu\nu}&=&\frac{N_{f}g^2}{2}\int\frac{d^3{\rm p}}{(2\p)^3}
\sum_{s=\pm}\frac{\theta_p^s}{2\veps_p^s}\nn\\
&\times& \left(\frac{K^2}{K^4 - 4(P.K)^2}
\sum_{s'=\pm}[{\cal M}_{\m\n}^{ss'}(P+K,P)
+{\cal M}_{\m\n}^{ss'}(P,P-K)]\right.\nn\\&&\left.
-\frac{2(P \cdot K)}{K^4 - 4(P.K)^2}
\sum_{s'=\pm} [{\cal M}_{\m\n}^{ss'}(P+K,P)
-{\cal M}_{\m\n}^{ss'}(P,P-K)]\right).
\eeq

\begin{figure}[htb]
\begin{center}
\resizebox{11.5cm}{5.0cm}{\includegraphics[]{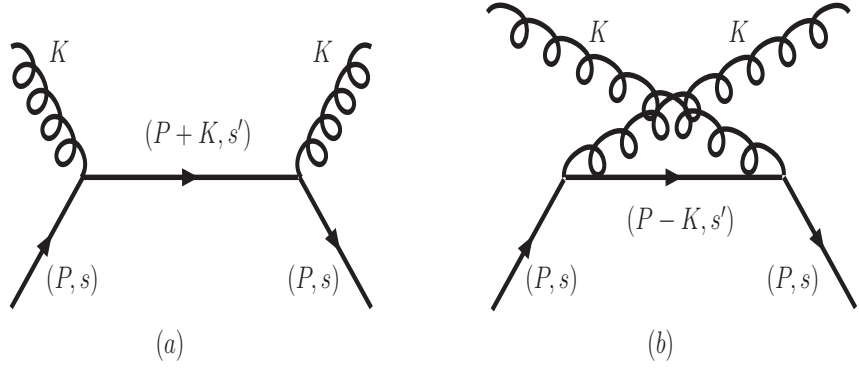}}
\caption{Compton scattering amplitude.}
\label{fig_02}
\end{center}
\end{figure}

Here, ${\cal M}_{\m\n}^{ss'}$ is related to the Compton 
scattering amplitude as shown in Fig.(\ref{fig_02}). To derive
Eq.(\ref{pi_munu}), following ref.\cite{tatsumi00,pal09}
we use projection operator
${\mathscr P}(a)=\frac{1}{2}(1+\gf a\sls)$
at each vertex. The momentum integration is performed at 
the Fermi surface restricted by $\theta^\pm_p = \theta(p^\pm_f -|p|)$.

Now we choose 
$K \equiv (k_{0},0,0,|k|)$, $P \equiv (\veps_{p},|p|{\sin\theta} {\cos\phi},
|p|{\sin\theta} {\sin\phi},|p|{\cos\theta})$, 
$s \equiv {\pm}({\sin\theta} {\cos\phi}, {\sin\theta} {\sin\phi}, 
{\cos\theta})$ and $ g_{\m\n}=(1,-1,-1,-1)$. 
Note that, the upper and lower cases fonts are used to distinguish
between four and three vectors.

From Fig.(\ref{fig_02}-a) scattering amplitude becomes

\beq\label{trace1}
{\cal M}_{\m\n}^{dir,ss'}(P+K,P)&=&
-b_{\mu}P_{\nu}(a.K)-g_{\m\n}(b.P)(a.K)+b_{\m}K_{\n}(a.P)
+a_{\m}P_{\n}(b.K)-g_{\m\n}(a.P)(b.K)\nn\\
&&+P_{\m}\{-K_{\n}(a.b-1)-2P_{\n}(a.b-1)+b_{\n}(a.K)+2b_{\n}(a.P)+a_{\n}(b.K)\}
\nn\\
&& +a_{\m}K_{\n}(b.P)+2a_{\m}P_{\n}(b.P)-2g_{\m\n}(a.P)(b.P)
-K_{\m}\{P_{\n}(a.b-1)-b_{\n}(a.P)\nn\\ && 
+a_{\n}(b.P)\}+(K.P)\{g_{\m\n}(a.b-1)-b_{\m}a_{\n}-a_{\m}b_{\n}\}.
\eeq

The components of the 4-pseudovector $b_{\mu}({\rm or~}a_{\m})$ in a 
frame in which the particle is moving with momentum $p({\rm or~}p+k)$
are found by the Lorentz transformation from the rest frame as given by 
{\cite{tatsumi00,pal09}},

\beq\label{vector1}
a_{0}&=&\frac{(p+k)\cdot s}{m_q};~~~
{\vec a}= s+\frac{(p+k)[(p+k)\cdot s]}{m_{q}(\veps_{p+k}+m_{q})};\nn\\
b_{0}&=&\frac{p\cdot s'}{m_q};~~~
{\vec b}= s'+\frac{p(p\cdot s')}{m_{q}(\veps_{p}+m_{q})}.
\eeq

Similarly, from Fig.(\ref{fig_02}-b) we have,

\beq\label{trace2}
{\cal M}_{\m\n}^{ex,ss'}(P,P-K)&=&-{\tilde b_{\mu}}P_{\nu}({\tilde a}.K)
+g_{\m\n}({\tilde b}.P)({\tilde a}.K)+{\tilde b}_{\m}K_{\n}({\tilde a}.P)
-{\tilde a}_{\m}P_{\n}({\tilde b}.K)+g_{\m\n}({\tilde a}.P)({\tilde b}.K)
\nn\\&&
-P_{\m}\{-K_{\n}({\tilde a}.{\tilde b}-1)+2P_{\n}({\tilde a}.{\tilde b}-1)
+{\tilde b}_{\n}({\tilde a}.K)-2{\tilde b}_{\n}({\tilde a}.P)
-{\tilde a}_{\n}({\tilde b}.K)\}
\nn\\&& 
-{\tilde a}_{\m}K_{\n}({\tilde b}.P)+2{\tilde a}_{\m}P_{\n}({\tilde b}.P)
-2g_{\m\n}({\tilde a}.P)({\tilde b}.P)
-K_{\m}\{-P_{\n}({\tilde a}.{\tilde b}-1)+{\tilde b}_{\n}({\tilde a}.P)
\nn\\ && 
+{\tilde a}_{\n}({\tilde b}.P)\}
-(K.P)\{g_{\m\n}({\tilde a}.{\tilde b}-1)-{\tilde b}_{\m}{\tilde a}_{\n}
-{\tilde a}_{\m}{\tilde b}_{\n}\},
\eeq

\noindent 

where,

\beq\label{vector2}
{\tilde a}_{0}&=&\frac{p\cdot s}{m_q};~~~
{\vec {\tilde a}}= s+\frac{p(p\cdot s)}{m_{q}(\veps_{p}+m_{q})}\nn\\
{\tilde b}_{0}&=&\frac{(p-k)\cdot s'}{m_q};~~~
{\vec {\tilde b}}= s'+\frac{(p-k)[(p-k)\cdot s']}{m_{q}(\veps_{p-k}+m_{q})}.
\eeq

Now we define matrix elements ${\cal M}_{\m\n}^{ss'}$ in 
terms of flip (f) and 
non-flip (nf) interaction where ${\cal M}_{\m\n}^{nf}=
{\cal M}_{\m\n}^{s=s'}$ 
and ${\cal M}_{\m\n}^{f}={\cal M}_{\m\n}^{s=-s'}$ \cite{tatsumi00,pal09}.
Using Eq.(\ref{trace1}) and Eq.(\ref{trace2}) we have,

\beq\label{long_tr}
{\cal M}_{00}^{f+nf}(P+K,P)+{\cal M}_{00}^{f+nf}(P,P-K)
&=&8\veps_{p}^2\nn\\
{\cal M}_{00}^{f+nf}(P+K,P)-{\cal M}_{00}^{f+nf}(P,P-K)
&=&-4(P.K)\nn\\
{\cal M}_{33}^{f+nf}(P+K,P)+{\cal M}_{33}^{f+nf}(P,P-K)
&=&8p^2{\cos^2{\theta}}\nn\\
{\cal M}_{33}^{f+nf}(P+K,P)-{\cal M}_{33}^{f+nf}(P,P-K)
&=&4[2pk{\cos\theta}+(P.K)]
\eeq

\noindent and 

\beq
{\cal M}_{11}^{f+nf}(P+K,P)+{\cal M}_{11}^{f+nf}(P,P-K)\nn\\
={\cal M}_{22}^{f+nf}(P+K,P)+{\cal M}_{22}^{f+nf}(P,P-K)
&=&4p^2{\sin^2{\theta}}\\
{\rm and~~~~~~~~~~~~}
{\cal M}_{11}^{f+nf}(P+K,P)-{\cal M}_{11}^{f+nf}(P,P-K)\nn\\
={\cal M}_{22}^{f+nf}(P+K,P)-{\cal M}_{22}^{f+nf}(P,P-K)&=&4(P.K).
\label{trans_tr}
\eeq

The detailed expressions of the matrix element ${\cal M}_{\m\n}^{ss'}$ 
are given in the Appendix. In the 
present work we consider one flavor quark matter. 
Generalization for multi-flavor system is straightforward. 
Using Eq.(\ref{pi_munu}) and Eq.(\ref{long_tr})-(\ref{trans_tr}) we get

\beq
\Pi_{11} &=&\frac{g^2}{8\p^3}\sum_{s=\pm}\int_{0}^{p_f^s}
\frac{d^3{\rm p}}{\veps_p}\left[\frac{K^2 p^2 \sin^2\theta-2(P\cdot K)^2}
{K^4-4(P\cdot K)^2}\right] \label{pi11} \\
\Pi_{00} &=&\frac{g^2}{4\p^3}\sum_{s=\pm}
\int_{0}^{p_f^s}\frac{d^3{\rm p}}{\veps_p}
\left[\frac{K^2 {\veps_p}^2 +(P\cdot K)^2}
{K^4-4(P\cdot K)^2}\right]\label{pi00}\\
\Pi_{33} &=&\frac{g^2}{4\p^3}\sum_{s=\pm}\int_{0}^{p_f^s}
\frac{d^3{\rm p}}{\veps_p}\left[\frac{K^2 p^2 \cos^2\theta
-2pk\cos\theta (P\cdot K) -(P\cdot K)^2}
{K^4-4(P\cdot K)^2}\right]\label{pi33}
\eeq

We are interested to evaluate longitudinal ($\Pi_L$) and 
transverse ($\Pi_T$) components of the polarization tensor. 
We define, $\Pi_L=-\Pi_{00}+\Pi_{33}$ and $\Pi_T=\Pi_{11}=\Pi_{22}$. 
In the long-wavelength limit ($|p| \sim p_f$ and $|k| \ll p_f$), 
{\rm i.e.} for low lying excitation near the Fermi surface, 
$K^4$ can be neglected compared to $4(P\cdot K)^2$ in the 
denominators of Eq.(\ref{pi11})-(\ref{pi33}) \cite{chin77}.
The longitudinal and transverse polarization in this limit 
are determined to be

\beq
\Pi_{L} &=&\frac{g^2}{4\p^2}(C_0^2-1)\sum_{s=\pm}
p_f^s\veps_f^s\left[-1+\frac{C_0}{2v_f^s}
\ln\left(\frac{C_0+v_f^s}{{C_0-v_f^s}}\right)\right],
\label{piL}\\
\Pi_{T}&=&\frac{g^2}{16\p^2}C_0\sum_{s=\pm}{p_f^s}^2
\left[\frac{2C_0}{v_f^s}+
\left(1-\frac{C_0^2}{{v_f^s}^2}\right)
\ln\left(\frac{C_0+v_f^s}{{C_0-v_f^s}}\right)\right]
\label{piT}.
\eeq

Here, we take $C_0=k_0/|k|$ and 
$v_f^{\pm}=p_f(1\pm\xi)^{1/3}/(p_f^2(1 \pm \xi)^{2/3}+m_q^2)^{1/2}$ in
order to cast the results in a more familiar form as presented
in \cite{chin77} for $\xi=0$. It might be noted here that, although the
final expressions for the longitudinal and transverse polarization
look rather similar to what one obtains in the case of unpolarized
matter \cite{chin77} with only difference in $v_f^\pm$
and summation over the spins, the calculation of the matrix elements
with explicit spin dependencies are rather involved (see Appendix). 

$\Pi_L$ and $\Pi_T$ have two limiting values, corresponding to the
non-relativistic (nr) and the ultra-relativistic (ur) regime.
In the non-relativistic limit 
$(\veps_f^{\pm} \rightarrow m_q)$

\beq
\Pi_{L}^{nr} &=&-\frac{g^2}{4\p^2}m_q\sum_{s=\pm}
p_f^s\left[-1+\frac{C_0}{2v_f^s}
\ln\left(\frac{C_0+v_f^s}{{C_0-v_f^s}}\right)\right],
\label{piL_nr}\\
\Pi_{T}^{nr}&=&\frac{g^2}{16\p^2}C_0\sum_{s=\pm}{p_f^s}^2
\left[\frac{2C_0}{v_f^s}+
\left(1-\frac{C_0^2}{{v_f^s}^2}\right)
\ln\left(\frac{C_0+v_f^s}{{C_0-v_f^s}}\right)\right]
\label{piT_nr}.
\eeq

Here $v_f^s=p_f^s/m_q$. 
These expressions were derived in \cite{wal_book,hols73} for 
unpolarized electron gas. In this limit, current-current interaction
is inherently small, for which this term can be neglected compared to 
the Coulomb interaction to calculate correlation energy. Here,
${\rm Re} \Pi_T \sim (k_0/|k|)^2$ and ${\rm Im} \Pi_T \sim (k_0/|k|)$ when both
$k_0 \rightarrow 0$ and $|k|\rightarrow 0$. It is apparent from this 
behavior of $\Pi_T$, that the current-current interaction remain 
unscreened at zero frequency \cite{hols73}.

In the ultra-relativistic limit $(\veps_f^s \rightarrow p_f^s)$  
the polarization tensors take the following forms

\beq
\Pi_L^{ur} &=& \frac{g^2}{4\p^2}\sum_{s=\pm}{p_f^s}^2 \sin^{-2}\theta_E
(1-\theta_E \cot\theta_E),
\label{piL_ur}\\
\Pi_T^{ur} &=& \frac{g^2}{8\p^2}\sum_{s=\pm}{p_f^s}^2 [1-\sin^{-2}\theta_E
(1-\theta_E \cot\theta_E)],
\label{piT_ur}
\eeq

with $\theta_E=\tan^{-1}(|k|/k_0)$. For $\xi=0$, these results are same as
those of ref.\cite{chin77}. In the next section, 
Eq.(\ref{piL})-(\ref{piT_ur}) are used to evaluate the contribution 
of the ring diagrams.

It might not be out of context here to mention that once we have the expressions
of $\Pi_{L(T)}$, one loop corrected gluon propagator in polarized quark matter
can easily be constructed. This forms the basis for calculation of various
physical quantities including the FLPs, which, without such
medium corrections, suffer from infrared divergences 
\cite{pal09,tatsumi08,baym76}.

\section{Ground state energy with correlation }

The leading contributions to the ground state energy are given by
the three terms {\em {viz.}} kinetic, exchange and correlation energy
densities {\rm i.e.}
 
\beq
E &=& E_{kin}+E_{ex}+E_{corr}+{\cal O}(r_s)
\eeq

In the high density limit ${\cal O}(r_s)$
vanish, the result becomes exact \cite{pines_book}.
$E_{kin}$ is given by \cite{tatsumi00,pal09}

\beq\label{ke}
E_{kin}&=&\frac{3}{16\p^2}
\left\{p_f(1+\xi)^{1/3}\sqrt{p_f^2(1+\xi)^{2/3}+m_q^2}
\left[2p_f^2(1+\xi)^{2/3}+m_q^2\right]\right.\nn\\&&\left.
-m_q^4\ln\left(\frac{p_f(1+\xi)^{1/3}+\sqrt{p_f^2(1+\xi)^{2/3}+m_q^2}}
{m_q}\right)
+[\xi\rightarrow -\xi]\right\},
\eeq

where $\xi$ is the polarization parameter
with the condition $0\le\xi\le 1$. Here $n_q^+$ and $n_q^-$ 
represent densities of spin-up and spin-down quarks
respectively and $n_{q}=n_{q}^{+}+n_{q}^{-}$ denote 
total quark density. Then the Fermi momenta in the spin-polarized
quark matter are defined as $p_{f}^{+}=p_{f}(1+{\xi})^{1/3}$ and 
$p_{f}^{-}=p_{f}(1-{\xi})^{1/3}$, where $p_{f}=(\pi^2n_{q})^{1/3}$, is the 
Fermi momentum of the unpolarized matter $({\xi}=0)$. 

In the non-relativistic (nr) and the ultra-relativistic (ur) limit kinetic 
energy density becomes \cite{tatsumi00,pal09},

\beq
E_{kin}^{nr}&=&\frac{3p_{f}^5}{20\pi^2m_{q}}
\left[(1+\xi)^{5/3}+(1-\xi)^{5/3}\right],
\label{kin_nr}\\
E_{kin}^{ur}&=&\frac{3p_{f}^4}{8\pi^2}
\left[(1+{\xi})^{4/3}+(1-{\xi})^{4/3}\right].
\label{kin_ur}
\eeq

\begin{figure}[htb]
\begin{center}
\resizebox{3.0cm}{4.4cm}{\includegraphics[]{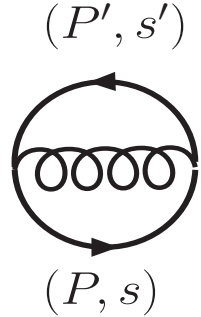}}
\caption{Two loop contribution to exchange energy density. Solid line
represents the quark propagator and the wavy line represents gluon.}
\label{fig_04}
\end{center}
\end{figure}

The first correction due to interaction to the ground state energy 
is given by the exchange energy density. 
This arises from two quarks interchanging positions 
in the Fermi sea by exchanging a virtual gluon \cite{freed77}. 
The exchange energy density was calculated in ref.\cite{pal09} within
Fermi liquid theory approach. One can directly 
evaluate the loop diagram to calculate $E_{ex}$ 
as shown in Fig.(\ref{fig_04}) \cite{tatsumi00}. 

For polarized quark matter, $E_{ex}$, consists of two terms 
$E_{ex}=E_{ex}^{nf}+E_{ex}^{f}$. Here {\cite{tatsumi00}},

\beq
E_{ex}^{nf}&=&\frac{9}{2}\sum_{s=\pm}\int\int\frac{d^3p}{(2\p)^3}
\frac{d^3p'}{(2\p)^3}\theta(p_f^s-|p|)\theta(p_f^s-|p'|)f_{pp'}^{nf},
\label{ex_nf}\\
E_{ex}^{f}&=& 9 \int\int\frac{d^3p}{(2\p)^3}
\frac{d^3p'}{(2\p)^3}\theta(p_f^{+}-|p|)\theta(p_f^{-}-|p'|)f_{pp'}^{f},
\label{ex_f}
\eeq

where $f_{pp'}$ is two particle forward scattering amplitude is given by
\cite{tatsumi00,pal09}

\beq
f_{pp'}^{ss'}&=&\frac{2g^2}{9\veps_p \veps_p'}\frac{1}{(P-P')^2}
[2m_{q}^2-P.P'-(p\cdot s)(p'\cdot s')
+m_{q}^2(s\cdot s')+\frac{1}{(\veps_{p}+m_{q})(\veps_{p'}+m_{q})}
\nn\\&&\times
\{m_{q}(\veps_{p}+m_{q})(p'\cdot s)(p'\cdot s')
+m_{q}(\veps_{p'}+m_{q})(p\cdot s)(p\cdot s')
+(p\cdot p')(p\cdot s)(p'\cdot s')\}].\nn\\
\eeq

where, $\veps_{p}=\sqrt{p^2+m_{q}^2}$. 
In the non-relativistic and the ultra-relativistic limit $E_{ex}$ yields
\cite{tatsumi00,pal09},

\beq
E_{ex}^{nr}&=&-\frac{g^2}{8\pi^4}p_{f}^4
\left[(1+\xi)^{4/3}+(1-\xi)^{4/3}\right],
\label{ex_nr}\\
E_{ex}^{ur}&=&\frac{g^2}{32\pi^4}p_{f}^4\left[(1+{\xi})^{4/3}+
(1-{\xi})^{4/3}+2(1-{\xi}^2)^{2/3}\right].
\label{ex_ur}
\eeq

\begin{figure}[htb]
\begin{center}
\resizebox{11.0cm}{2.0cm}{\includegraphics[]{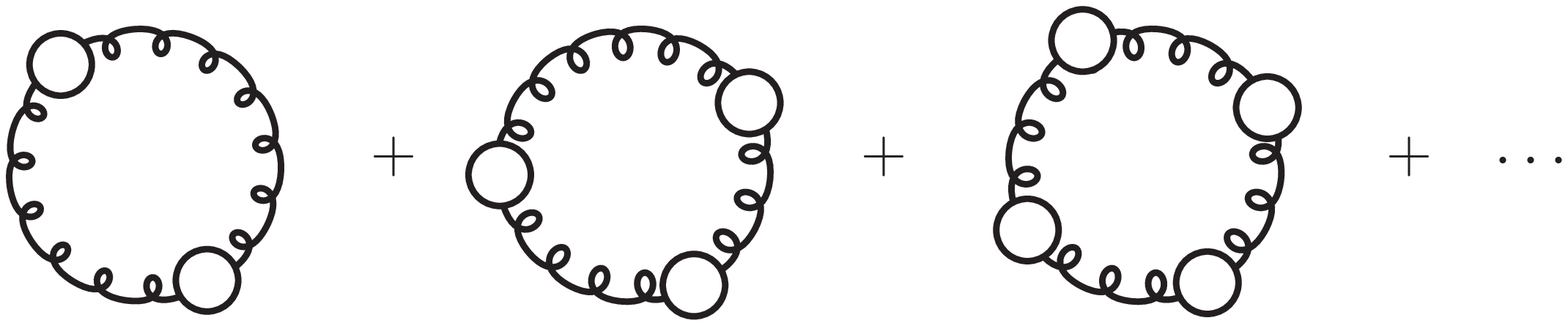}}
\caption{The series of ring diagrams.}
\label{ring_dia}
\end{center}
\end{figure}


Now we come to the central aim of the present work, {\em i.e.} the
evaluation of the correlation energy of dense quark matter with
arbitrary spin polarization; the leading contribution to $E_{corr}$
can be obtained by adding the contributions of ring diagrams as shown in
Fig.\ref{ring_dia}. It is to be noted that each of these diagrams are 
infrared divergent while their sum is finite 
\cite{gellmann57,pines58,chin77,sawada57,freed77} and are given by:

\beq\label{ring1}
E_{corr}&=&E_{corr}^L+E_{corr}^T\nn\\
&=&-\frac{i}{2}\int\frac{d^4{\rm K}}{(2\p)^4}
\Big([\ln(1-D^0\Pi_L)+D^0\Pi_L]+2[\ln(1-D^0\Pi_T)+D^0\Pi_T]\Big).
\eeq

Here $D^0$ is the free gluon propagator. The spatial integral of 
Eq.(\ref{ring1}) can be reduced to one for the radial variable only, 
because all the polarization propagators are independent of the 
direction of three momentum transfer $k$. A Wick rotation is 
performed on the fourth component of the integration momentum 
$(k_0 \rightarrow ik_0)$ so that space metric becomes Euclidean
\cite{chin77,ji88}. With $K_E^2=k_0^2+|k|^2=-K^2$ and $\tan\theta_E=|k|/k_0$,
Eq.(\ref{ring1}) becomes,

\beq\label{ring2}
E_{corr} &=& \frac{1}{(2\p)^3}\int_0^{\infty} K_E^2 {\rm d}K_E^2
\int_0^{\p/2}\sin^2\theta_E {\rm d}\theta_E \left(
\left[\ln\left(1+\frac{\Pi_L(K_E^2,\theta_E)}{K_E^2}\right)
-\frac{\Pi_L(K_E^2,\theta_E)}{K_E^2}\right]\right.\nn\\
&&\left.+2\left[\ln\left(1+\frac{\Pi_T(K_E^2,\theta_E)}{K_E^2}\right)
-\frac{\Pi_T(K_E^2,\theta_E)}{K_E^2}\right]\right)
\eeq

Infrared divergences would arise in Eq.(\ref{ring2}), if we were
to expand the logarithms in powers of $\Pi_i$ because of the non-zero value 
of $\Pi_i(K_E^2,\theta_E)$ at $K_E^2=0$. This can be isolated by writing 
$K_E^2=0$ whenever possible in the integrand.
Following ref.\cite{freed77,kap_book}, we have

\beq\label{ring3}
E_{corr} & \simeq & \frac{1}{(2\p)^3}\int_0^{\infty} K_E^2 {\rm d}K_E^2
\int_0^{\p/2}\sin^2\theta_E {\rm d}\theta_E
\Big\{\left[\ln\left(1+\frac{\Pi_L(0,\theta_E)}{K_E^2}\right)
-\frac{\Pi_L(0,\theta_E)}{K_E^2}\right]\nn\\
&+&2\left[\ln\left(1+\frac{\Pi_T(0,\theta_E)}{K_E^2}\right)
-\frac{\Pi_T(0,\theta_E)}{K_E^2}\right]
+\frac{1}{2K_E^2}\frac{1}{K_E^2+\veps_f^2}
[\Pi_L^2(0,\theta_E)+2\Pi_T^2(0,\theta_E)]\Big\}\nn\\
\eeq

Performing $K_E^2$ integration the ring energy becomes 
\cite{chin77,freed77,kap_book}

\beq\label{ring4}
E_{corr} & \simeq & \frac{1}{(2\p)^3}\frac{1}{2}
\int_0^{\p/2}\sin^2\theta_E {\rm d}\theta_E
\left(\Pi_L^2\left[\ln\left(\frac{\Pi_L}{\veps_f^2}\right)-\frac{1}{2}\right]
+2\Pi_T^2\left[\ln\left(\frac{\Pi_T}{\veps_f^2}\right)-\frac{1}{2}\right]
\right)
\eeq

To proceed further, we first express $\Pi_L$ and $\Pi_T$
in terms of polar variables. From Eq.(\ref{piL}) and Eq.(\ref{piT})
we obtain

\beq
\Pi_L &=& \frac{g^2}{4\p^2}\sum_{s=\pm}\frac{p_f^s\veps_f^s}{\sin^2\theta_E}
\left[1-\frac{\cot\theta_E}{v_f^s}
\tan^{-1}\left(v_f^s\tan\theta_E \right)\right],
\label{piL_polar}\\
\Pi_T &=& \frac{g^2}{8\p^2}\sum_{s=\pm}{p_f^s}^2\cot\theta_E
\left[-\frac{\cot\theta_E}{v_f^s}+
\left(1+\frac{\cot^2\theta_E}{{v_f^s}^2}\right)
\tan^{-1}\left(v_f^s\tan\theta_E \right)\right].
\label{piT_polar}
\eeq

\begin{figure}[htb]
\begin{center}
\resizebox{9.5cm}{8.0cm}{\includegraphics[]{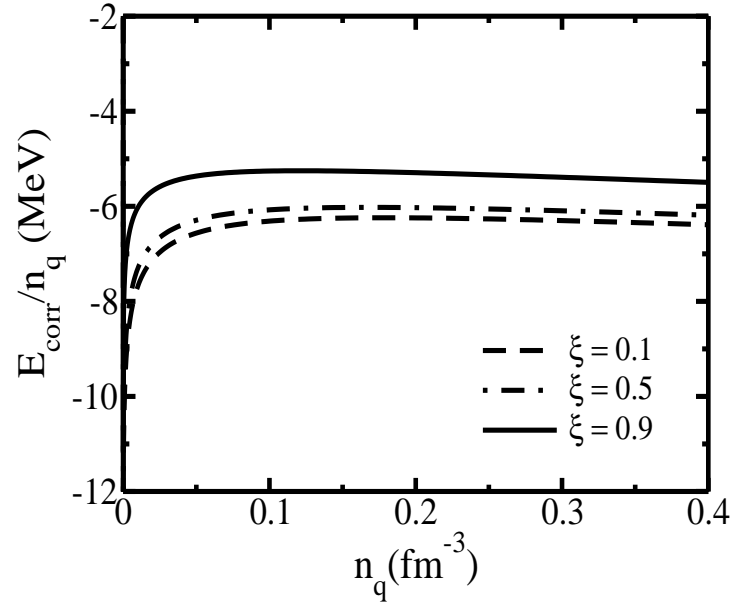}}
\caption{Correlation energy $(E_{corr})$ as a function of density 
for different polarization parameters.}
\label{corr_eng}
\end{center}
\end{figure}

These are then inserted in Eq.(\ref{ring4}) and $\theta_E$ integration 
is performed numerically to estimate $E_{corr}$ for various $\xi$
as shown in Fig.(\ref{corr_eng}).

\vskip 0.2in


We can also derive the analytic expression for the correlation
energy in the non-relativistic and ultra-relativistic case by using 
relevant $\Pi_L$ and $\Pi_T$ as given in 
Eq.(\ref{piL_nr})-(\ref{piT_nr}) and Eq.(\ref{piL_ur})-(\ref{piT_ur})
respectively.

In the non-relativistic limit it is given by 

\beq\label{corr_nr}
E_{corr}^{nr}&=& \frac{g^4\ln g^2}{(2\p)^6}(1-\ln 2)\frac{1}{3}m_q p_f^3 
\eeq

Note that the correlation energy here is independent of spin polarization
parameter $\xi$. This is because it is proportional to $p_f^3$ when
$\xi$ dependent terms cancel. In deriving Eq.(\ref{corr_nr}) 
we consider exchange of longitudinal gluons only. 
It is to be mentioned that similar expressions for degenerate 
electron gas interacting via. static Coulomb potential can be found in 
ref.\cite{kap_book,akhi60}.

In the ultra-relativistic limit, the leading $g^4\ln g^2$ order contribution 
to $E_{corr}^L$ is derived to be

\beq\label{corr_urL}
E_{corr}^{ur, L} &=& \frac{g^4\ln g^2}{(2\p)^6}(1-\ln 2)
\frac{1}{12}p_f^4 [(1+\xi)^{4/3}+(1-\xi)^{4/3}+2(1-\xi^2)^{2/3}]
\eeq 

The term $(1-\ln 2)$ is the reminiscent of what one obtains in 
the non-relativistic electron plasma as was first obtained by
GB \cite{gellmann57}. In the relativistic case, such a term does 
not appear in the final expression of $E_{corr}$, where
a similar term with opposite sign arise out of the 
magnetic interaction mediated by the exchange of transverse gluons as :

\beq\label{corr_urT}
E_{corr}^{ur, T} &=& \frac{g^4\ln g^2}{(2\p)^6}(\ln 2 -\frac{5}{8})
\frac{1}{12}p_f^4 [(1+\xi)^{4/3}+(1-\xi)^{4/3}+2(1-\xi^2)^{2/3}]
\eeq
 
By adding Eq.(\ref{corr_urL}) and Eq.(\ref{corr_urT}) one obtains

\beq\label{corr_ur}
E_{corr}^{ur} &=& \frac{g^4\ln g^2}{2048 \p^6}p_f^4
[(1+\xi)^{4/3}+(1-\xi)^{4/3}+2(1-\xi^2)^{2/3}].
\eeq

For $\xi=0$, the correlation energy for unpolarized matter follows
\cite{chin77,freed77,akhi60}. That the term $\ln 2$ disappear from
the relativistic ring energy is known from the work \cite{chin77}
where a detailed calculation of the correlation energy for the nuclear
matter ground state has been performed. Furthermore, one may also
note that in the non-relativistic limit $E_{ex}$ and $E_{corr}$ 
contribute with opposite sign while in the ultra-relativistic limit, 
both of them contribute with same sign.

\vskip 0.2in
\begin{figure}[htb]
\begin{center}
\resizebox{9.5cm}{8.0cm}{\includegraphics[]{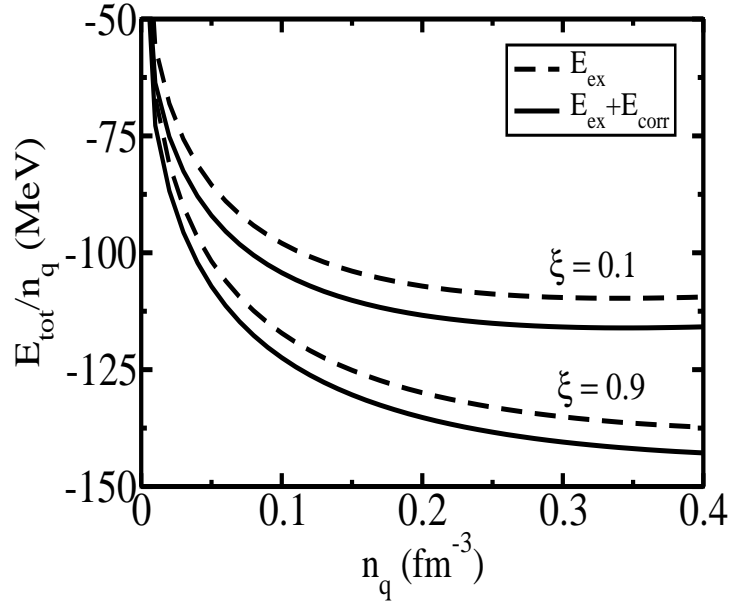}}
\caption{Comparison of exchange and correlation energy in 
polarized quark matter as a function of density 
for different polarization parameters.}
\label{ex_corr}
\end{center}
\end{figure}

\begin{figure}[htb]
\begin{center}
\resizebox{9.5cm}{8.0cm}{\includegraphics[]{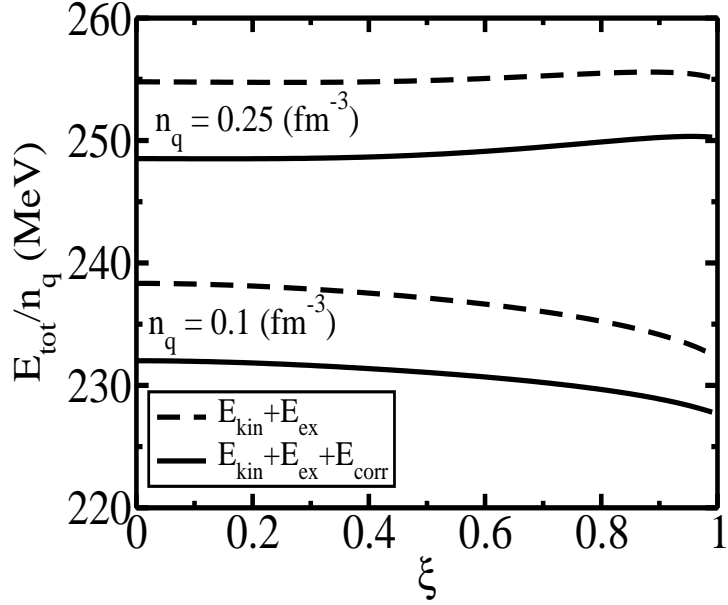}}
\caption{Total energy of quark liquid as a function of polarization
parameter at $n_q=0.1 {\rm fm^{-3}}$ and $n_q=0.25 {\rm fm^{-3}}$. 
The critical density is around $n_q^c=0.16 {\rm fm^{-3}}$ in this case.}
\label{phase}
\end{center}
\end{figure}

Using Eq.(\ref{piL_polar}) and Eq.(\ref{piT_polar}) correlation 
energy is estimated numerically which is valid for all the kinematic regimes. 
For this, following ref.\cite{tatsumi00,pal09}, 
we take $\alpha_{c}=g^2/{4\pi}=2.2$ and $m_{q}=300 MeV$. 
In Fig.(\ref{corr_eng}) we plot density dependence of correlation energy 
for various $\xi$. This shows that at a given density, with higher value 
of $\xi$, $E_{corr}$ increases. 
In Fig.(\ref{ex_corr}), we compare 
exchange and correlation energy density. It shows system
becomes more bound when quark matter changes its phase from unpolarized to 
polarized matter. With increasing $\xi$, $E_{corr}$ remains attractive, 
however, its value decreases as observed both in Fig.(\ref{corr_eng}) and 
(\ref{ex_corr}). In Fig.(\ref{phase}) we plot ground state energy as 
a function of polarization parameter $\xi$. 
Hence we conclude that the quark matter interacting via. OGE 
becomes polarized at lower density while at higher density its 
becomes unpolarized. 
This clearly shows phase transition is first order and critical density 
is still around normal nuclear matter density 
$n _q^c \sim 0.16 {\rm fm^{-3}}$ \cite{tatsumi00,pal09}. In this regime,
it is seen, that $E_{corr}$ makes the system more bound.

To derive the spin stiffness constant in the high density limit 
using Eq.(\ref{kin_ur}), (\ref{ex_ur}) and (\ref{corr_ur}) we have,

\beq
\beta_s &=& \frac{\del^2 E}{\del\xi^2}{\Big|}_{\xi=0}\nn\\
&=& \beta_s^{kin}+\beta_s^{ex}+\beta_s^{corr}\nn\\
&=&\frac{p_f^4}{3\pi^2}\left[
1-\frac{g^2}{6\p^2}-\frac{g^4}{384\p^4}(\ln r_s-0.286)\right].
\eeq

Here, the logarithmic term arises from the correlation correction.

\vskip 0.2in 


\section{Summary and Conclusion}

In this work we derive the expressions for the gluon-self energy in
spin polarized quark matter and calculate the ground state energy of such a 
system upto term ${\cal O}(g^4)$ which include
corrections due to correlation effects. The analytical expressions for
the correlation energy in two limiting cases (non/ultra-relativistic) 
are presented and compared with $E_{ex}$ and $E_{kin}$. It is shown that the
correlation energy for polarized quark matter is comparatively larger 
than the unpolarized one, although it is always attractive. We find that 
numerically
the contribution of $E_{corr}$ to the total energy is not found to be large
and therefore, although qualitatively important, it is not the main factor in 
determining whether quark matter is ferromagnetic or not. 
With out this, however, the results remain 
incomplete because of the associated divergences of the terms beyond 
exchange diagrams \cite{wal_book,pines_book}. Furthermore, this is an
important first step to include the corrections due to correlations to 
the spin-susceptibility \cite{brueck58, shastry77, her_book}. 
In this work we present spin stiffness constant $\beta_s$ of dense 
quark system only in the high density limit. A detailed study of this
is now underway and shall be reported elsewhere\cite{pal_prep}. 

The inclusion of correlation energy, as shown here, 
does not rule out the possibility of ferromagnetic phase transition in 
quark matter at low density, rather, makes it more probable within the
model and parameter set used by Tatsumi \cite{tatsumi00} which
was borrowed from the bag model and was also used in \cite{chin79prl}.
Clearly the critical density at which the spin-polarized ferromagnetic state
might appear depends strongly on the quark mass and the critical density 
increases with increasing mass, this might change our numerical estimates.

Further uncertainty to the estimation of the critical density from the 
present analysis comes from the fact that we here
restrict ourselves only to OGE diagrams and one flavor system. 
In this regime, multi-gluon exchange processes {\cite{ohnishi07}} might play 
an important role. More work in this direction is therefore necessary
to examine this issue especially for multi flavor system which might appear in 
astrophysics. Leaving aside these questions, the evaluation of the gluon 
self-energy and the estimation of correlation energy 
in polarized matter, as mentioned in the text, nevertheless constitute an 
important component for the study of the properties of dense quark system.

\vskip 0.2in
{\bf Acknowledgments}\\

The authors would like to thank Samir Mallik for 
the critical reading of the manuscript.


\section{Appendix}

In the text Compton scattering amplitudes are given as a sum of  
flip and non-flip terms.
Here we give detail expression of 
${\cal M}_{11}^{ss'}(P+K,P)+{\cal M}_{11}^{ss'}(P,P-K)$ 
with explicit spin indices. With the help of Eq.(\ref{trace1}) and 
Eq.(\ref{trace2}) we have,

\beq
{\cal M}_{11}^{ss'}(P+K,P)+{\cal M}_{11}^{ss'}(P,P-K)&=&
\mathscr{A}_1+\mathscr{A}_2+\mathscr{A}_3
+\mathscr{A}_4+\mathscr{A}_5+\mathscr{A}_6
\eeq
where
\beq
\mathscr{A}_1&=&(b\cdot P)(a\cdot K)-({\tilde b}\cdot P)({\tilde a}\cdot K)\nn\\
&=&\frac{\veps_p k_0}{m_{q}^2}[(p\cdot s)(k\cdot s')+(k\cdot s)(p\cdot s')]
-\frac{k_0}{m_q}[(k\cdot s)(p\cdot s')]\nn\\
&&-\frac{k_0 p^2}{m_{q}^2(\veps_p+m_q)}[(p\cdot s)(k\cdot s')+(k\cdot s)(p\cdot s')]
-\frac{\veps_p}{m_{q}}[(k\cdot s)(k\cdot s')]\nn\\
&&+\frac{p^2}{m_{q}(\veps_p+m_q)}[(k\cdot s)(k\cdot s')]
-\frac{\veps_p (p\cdot k)}{m_{q}^2(\veps_p+m_q)}
[(p\cdot s)(k\cdot s')+(k\cdot s)(p\cdot s')]\nn\\
&&+\frac{(p\cdot k)}{m_{q}(\veps_p+m_q)}
[2(k\cdot s)(p\cdot s')-(k\cdot s)(k\cdot s')]\nn\\
&&-\frac{k_0(p\cdot k)}{m_{q}^2(\veps_p+m_q)}
[(p\cdot s)(p\cdot s')-(p\cdot s)(k\cdot s')]\nn\\
&&+\frac{p^2(p\cdot k)}{m_{q}^2(\veps_p+m_q)^2}
[(p\cdot s)(k\cdot s')+(k\cdot s)(p\cdot s')]\nn\\
&&+\frac{p^2 k^2}{m_{q}^2(\veps_p+m_q)^2}
[(p\cdot s)(p\cdot s')+(k\cdot s)(p\cdot s')]\nn\\
&&+\frac{(p\cdot k)^2}{m_{q}^2(\veps_p+m_q)^2}
[(p\cdot s)(p\cdot s')-(p\cdot s)(k\cdot s')]\nn\\
&&-\frac{\veps_p k^2}{m_{q}^2(\veps_p+m_q)}
[(p\cdot s)(p\cdot s')+(k\cdot s)(p\cdot s')]\nn\\
&&+\frac{k^2}{m_{q}(\veps_p+m_q)}
[(p\cdot s)(p\cdot s')+(k\cdot s)(p\cdot s')],
\eeq

\beq
\mathscr{A}_2&=&(a\cdot P)(b\cdot K)-({\tilde a}\cdot P)({\tilde b}\cdot K)\nn\\
&=&\frac{\veps_p k_0}{m_{q}^2}[(p\cdot s')(k\cdot s)+(p\cdot s)(k\cdot s')]
-\frac{\veps_p}{m_{q}}[(k\cdot s)(k\cdot s')]\nn\\
&&-\frac{k_0}{m_{q}}[(p\cdot s)(k\cdot s')]
+\frac{\veps_p (p\cdot k)}{m_{q}^2(\veps_p+m_q)}
[(p\cdot s)(k\cdot s')+(k\cdot s)(p\cdot s')]\nn\\
&&+\frac{p\cdot k}{m_{q}(\veps_p+m_q)}
[2(p\cdot s)(k\cdot s')+(k\cdot s)(k\cdot s')]
+\frac{p^2}{m_{q}(\veps_p+m_q)}[(k\cdot s)(k\cdot s')]\nn\\
&&-\frac{p^2 k_0}{m_{q}^2(\veps_p+m_q)}
[(p\cdot s)(k\cdot s')+(k\cdot s)(p\cdot s')]\nn\\
&&+\frac{p^2(p\cdot k)}{m_{q}^2(\veps_p+m_q)^2}
[(p\cdot s)(k\cdot s')+(k\cdot s)(p\cdot s')]\nn\\
&&-\frac{k_0(p\cdot k)}{m_{q}^2(\veps_p+m_q)}
[(p\cdot s)(p\cdot s')+(k\cdot s)(p\cdot s')]\nn\\
&&+\frac{(p\cdot k)^2}{m_{q}^2(\veps_p+m_q)^2}
[(p\cdot s)(p\cdot s')+(k\cdot s)(p\cdot s')]\nn\\
&&-\frac{k^2\veps_p}{m_{q}^2(\veps_p+m_q)}
[(p\cdot s)(p\cdot s')-(p\cdot s)(k\cdot s')]\nn\\
&&+\frac{k^2}{m_{q}(\veps_p+m_q)}
[(p\cdot s)(p\cdot s')-(p\cdot s)(k\cdot s')]\nn\\
&&+\frac{p^2 k^2}{m_{q}^2(\veps_p+m_q)^2}
[(p\cdot s)(p\cdot s')-(p\cdot s)(k\cdot s')]
\eeq

\beq
\mathscr{A}_3&=&2{p_1}^2[(a\cdot b)+({\tilde a}\cdot{\tilde b})-2]\nn\\
&=&p^2{\sin^2\theta}\left[\frac{1}{{m_q}^2}
[2(p\cdot s)(p\cdot s')+(p\cdot s')(k\cdot s)
-(p\cdot s)(k\cdot s')]-2(s\cdot s') \right.\nn\\&& \left.
-\frac{1}{m_{q}(\veps_p+m_q)}
[4(p\cdot s)(p\cdot s')+2(k\cdot s)(k\cdot s')]\right.\nn\\&& \left.
-\frac{p^2}{m_{q}^2(\veps_p+m_q)^2}
[2(p\cdot s)(p\cdot s')-(p\cdot s)(k\cdot s')+(p\cdot s')(k\cdot s)]
\right.\nn\\&& \left.
-\frac{(p\cdot k)}{m_{q}^2(\veps_p+m_q)^2}
[(p\cdot s)(k\cdot s')+(p\cdot s')(k\cdot s)]-2\right]
\eeq

\beq
\mathscr{A}_4&=&2[(a\cdot P)(b\cdot P)+({\tilde a}\cdot P)({\tilde b}\cdot P)]\nn\\
&=&2\left[\frac{{\veps_p}^2}{m_{q}^2}
[2(p\cdot s)(p\cdot s')+(p\cdot s')(k\cdot s)-(p\cdot s)(k\cdot s')]
\right. \nn\\ && \left.
-\frac{\veps_p}{m_{q}}
[4(p\cdot s)(p\cdot s')+(p\cdot s')(k\cdot s)-(p\cdot s)(k\cdot s')]
\right.\nn\\&& \left.
+\frac{p^2}{m_{q}(\veps_p+m_q)}
[4(p\cdot s)(p\cdot s')+(p\cdot s')(k\cdot s)-(p\cdot s)(k\cdot s')]
\right.\nn\\&& \left.
-\frac{\veps_p p^2}{m_{q}^2(\veps_p+m_q)}
[4(p\cdot s)(p\cdot s')+2(p\cdot s')(k\cdot s)-2(p\cdot s)(k\cdot s')]
\right.\nn\\&& \left.
+\frac{p^2(p\cdot k)}{m_{q}^2(\veps_p+m_q)^2}
[(p\cdot s)(k\cdot s')+(p\cdot s')(k\cdot s)]+2(p\cdot s)(p\cdot s')
\right.\nn\\&& \left.
-\frac{\veps_p (p\cdot k)}{m_{q}^2(\veps_p+m_q)}
[(p\cdot s)(k\cdot s')+(p\cdot s')(k\cdot s)]
\right.\nn\\&& \left.
+\frac{(p\cdot k)}{m_{q}(\veps_p+m_q)}
[(p\cdot s)(k\cdot s')+(p\cdot s')(k\cdot s)]
\right.\nn\\&& \left.
+\frac{p^4}{m_{q}^2(\veps_p+m_q)^2}
[2(p\cdot s)(p\cdot s')-(p\cdot s)(k\cdot s')+(p\cdot s')(k\cdot s)]
\right]
\eeq

\beq
\mathscr{A}_5&=&2(P\cdot K)(a_1 b_1-{\tilde a_1}{\tilde b_1})\nn\\
&=&\frac{p^2{\sin^2\theta}(P\cdot K)}{m_{q}^2(\veps_p+m_q)^2}
\left[(p\cdot s)(k\cdot s')+(p\cdot s')(k\cdot s)\right]
\eeq

\beq
\mathscr{A}_6&=&(P\cdot K)(a\cdot b-{\tilde a}\cdot{\tilde b})\nn\\
&=&\frac{(P\cdot K)}{m_{q}^2}
\left[[(p\cdot s)(k\cdot s')+(p\cdot s')(k\cdot s)]
\right.\nn\\&& \left.
-\frac{2m_q}{(\veps_p+m_q)}
[(p\cdot s)(k\cdot s')+(p\cdot s')(k\cdot s)]
\right.\nn\\&& \left.
-\frac{p^2}{(\veps_p+m_q)^2}
[(p\cdot s)(k\cdot s')(p\cdot s')(k\cdot s)]
\right.\nn\\&& \left.
+\frac{p\cdot k}{(\veps_p+m_q)^2}
[2(p\cdot s)(p\cdot s')-(p\cdot s)(k\cdot s')+(p\cdot s')(k\cdot s)]
\right]
\eeq

Similarly, one can derive terms like 
$[{\cal M}_{11}(P+K,P)-{\cal M}_{11}(P,P-K)]$, 
$[{\cal M}_{22}(P+K,P){\pm}{\cal M}_{22}(P,P-K)]$ etc. with the help of 
Eq.(\ref{vector1}) and Eq.(\ref{vector2}). After explicit calculation of 
those terms, $\Pi_{L,T}$ can be evaluate.


\begin{thebibliography}{99}
\bibitem{tatsumi00} T.Tatsumi, Phys.Lett.{\bf B 489}, 280 (2000).
\bibitem{niegawa05} A.Niegawa, Prog.Theor.Phys {\bf 113}, 581 (2005).
\bibitem{tatsumi07} T.Tatsumi, arXiv:0711.3349 (2007).\\
    T.Tatsumi, arXiv: astro-ph/0004062, (2000).
\bibitem{pal09} K.Pal, S.Biswas and A.K.Dutt-Mazumder, Phys.Rev.{\bf C79},
015205 (2009).
\bibitem{ohnishi07} K.Ohnishi, M.Oka and S.Yasui, Phys.Rev.{\bf D76}, 
097501 (2007).
\bibitem{son08} D.T.Son and M.A.Stephanov, Phys.Rev.{\bf D77}, 014021 (2008).
\bibitem{gellmann57} M.Gell-Mann and K.A.Brueckner, 
Phys.Rev.{\bf 106}, 364 (1957).
\bibitem{pines58} P.Nozi\'{e}res and D.Pines, 
Phys.Rev.{\bf 111}, 442 (1958).
\bibitem{bloch29} F.Bloch, Z.Phys.{\bf 57}, 545 (1929).
\bibitem{chin79prl} S.A.Chin and A.K.Kerman, Phys.Rev Lett. {\bf 43},
1292 (1979).
\bibitem{tatsumi08} T.Tatsumi and K.Sato, Phys.Lett.{\bf B663}, 322 (2008).
\bibitem{wal_book} A.L.Fetter and J.D.Walecka, {\it Quantum theory of 
many-particle systems}, (Dover Publications, INC. New York.)
\bibitem{pines_book} D.Pines, {\it The many-body problem}, 
Lecture Note, (New York, 1961).
\bibitem{perdew92} J.P.Perdew and Y.Wang, Phys.Rev.{\bf B45}, 13244 (1992).
\bibitem{rajiv89} R.R.P.Singh and D.A.Huse, Phys.Rev.{\bf B40}, 7247 (1989).
\bibitem{baym76} G.Baym and S.A.Chin, Nucl Phys.{\bf A262}, 527 (1976).
\bibitem{matsui81} T.Matsui, Nucl.Phys.{\bf A370}, 365 (1981).
\bibitem{dumitru09} A.Dumitru, Y.Guo and M.Strickland, arXiv:0903.4703.
\bibitem{bel_book} M.L.Bellac, {\it Thermal field theory},
(Cambridge University Press), (1996).
\bibitem{niegawa03} A.Niegawa, Phys.Rev.{\bf D68}, 116007 (2003).
\bibitem{chin77} S.A.Chin, Ann.Phys. {\bf 108}, 301 (1977).
\bibitem{hols73} T.Holstein, R.E.Norton and P.Pincus, 
Phys.Rev.{\bf B 8}, 2649 (1973).
\bibitem{freed77} B.A.Freedman and L.D.McLerran, Phys.Rev.{\bf D 16},
 1130 (1977).\\
B.A.Freedman and L.D.McLerran, Phys.Rev.{\bf D 16}, 1147 (1977).\\
B.A.Freedman and L.D.McLerran, Phys.Rev.{\bf D 16}, 1169 (1977).
\bibitem{sawada57} K.Sawada, K.A.Brueckner and N.Fukuda,
Phys.Rev.{\bf 108}, 507 (1957).
\bibitem{ji88} X.Ji, Phys.Lett.{\bf B 208}, 19 (1988).
\bibitem{kap_book} J.I.Kapusta, {\it Finite temperature Field theory},
(Cambridge University Press), (1989).
\bibitem{akhi60} I.A.Akhiezer and S.V.Peletminskii, 
{\it Soviet Physics} {\bf JETP 11}, 1316 (1960).
\bibitem{brueck58} K.A.Brueckner and K.Sawada, Phys.Rev.{\bf 112}, 328 (1958).
\bibitem{shastry77} B.S.Shastry, Phys.Rev.Lett.{\bf 38}, 449 (1977).
\bibitem{her_book} C.Herring, {\it Magnetism}, edited by G.T.Rado and H.Shul
 (Academic, New York, 1966), Vol. IV.
\bibitem{pal_prep} K.Pal and A.K.Dutt-Mazumder, (in preparation).
\end{thebibliography}
\end{document}